\documentclass[a4paper,11pt]{article}
\pdfoutput=1 % if your are submitting a pdflatex (i.e. if you have
\usepackage{jinstpub} % for details on the use of the package, please
\usepackage{subcaption}
\title{Characterisation of SiPM radiation hardness for application in hadron calorimeters at FAIR, CERN and NICA}

\author[a,b,1]{V.~Mikhaylov,\note{Corresponding author.}}
\author[c,d]{F.~Guber,}
\author[c,d]{A.~Ivashkin,}
\author[a]{A.~Kugler,}
\author[a]{V.~Kushpil,}
\author[c,e]{S.~Morozov,}
\author[a]{O.~Svoboda,}
\author[a]{and P.~Tlust\'{y}}

\affiliation[a]{Nuclear Physics Institute, Czech Academy of Sciences, Hlavn\'{i} 130, 25068 Husinec - \v{R}e\v{z}, Czech Republic}
\affiliation[b]{Czech Technical University in Prague, Faculty of Nuclear Sciences and Physical Engineering, B\v{r}ehov\'{a} 7, 11519 Prague, Czech Republic}
\affiliation[c]{Institute for Nuclear Research, Russian Academy of Sciences, Prospect 60-letiya Octyabrya 7-a, 117312 Moscow, Russia}
\affiliation[d]{Moscow Institute of Physics and
Technology, 1 "A" Kerchenskaya st., 117303 Moscow, Russia}
\affiliation[e]{National Research Nuclear University MEPhI, Kashirskoe shosse 31, 115409 Moscow, Russia}

\emailAdd{mikhaylov@ujf.cas.cz}

\abstract{Silicon PhotoMultipliers (SiPM) are an excellent choice for the scintillator light readout at hadron calorimeters due to their insensitivity to magnetic fields, low operating voltages, low cost, compactness and mechanical endurance. They are already successfully utilized in Projectile Spectator Detector (PSD) of NA61 at CERN, and will be utilized soon in PSD of CBM at FAIR and Forward Hadron CALorimeters (FHCAL) of BM@N at NICA heavy-ion collision experiments. The main issue of SiPM application is their degradation due to high neutron fluence that can reach up to $2\times10^{11}$~n$_{\textrm{eq}}$/cm$^2$ per year of the experiment operation. Multiple irradiation tests of SiPMs produced by Ketek, Zecotek, Hamamatsu and Sensl manufacturers were conducted at the cyclotron of NPI \v{R}e\v{z} with a broad neutron spectrum and total fluences in the wide range of $5\times10^{10}$ -- $6\times10^{12}$ n$_{\textrm{eq}}$/cm$^2$. Detailed characterisation  of all SiPMs was performed based on dependencies of dark current on voltage, capacitance on voltage and frequency, and response to LED light on voltage. SiPM's breakdown voltage, quenching resistance, pixel capacitance, gain and signal to noise ratio were extracted from these measurements. Those parameters' dependence on neutron fluence and their variability are discussed. Performance of the PSD calorimeter module equipped with irradiated SiPMs in CERN during the beam scan with 2 – 80 GeV/c protons is briefly overviewed.}


\proceeding{21$^{\text{th}}$  International Workshop on Radiation Imaging Detectors\\
  7–12 July 2019\\
  Kolymbari, Crete, Greece}

\begin{document}
\maketitle
\flushbottom

\section{Introduction}
\label{sec:intro}

NA61@CERN, CBM@FAIR, BM@N and NICA heavy-ion collision experiments employ compensating lead-scintillator calorimeters to measure the energy distribution of the forward going projectile nucleons and nuclei fragments (spectators) produced close to the beam rapidity~\cite{GuberNew}. 
%They provide information on the centrality and reaction plane orientation of the collision. 
%These detectors are called Projectile Spectator Detectors (PSD) at NA61 at CERN and CBM at FAIR, Forward Hadron CALorimeters (FHCALs) and Zero Degree Calorimeters (ZDC) at BM@N and MPD at NICA. 
The scintillation light is transferred via the WaveLength Shifting fibers and read out by Silicon PhotoMultipliers (SiPM). High interaction rates up to 1 MHz lead to harsh radiation conditions, namely total ionization dose up to 1 kGy and neutron fluence up to $2\times10^{11}$~n$_{\textrm{eq}}$/cm$^2$ per year of the experiment operation. Only negligible changes in scintillators' light yield were observed after irradiation with this ionization dose~\cite{Guber:109059}. In this article we compare changes of main parameters for various SiPMs after the neutron irradiation. The list of investigated SiPMs and their parameters, namely breakdown voltage $V_{bd}$, number of pixels $N_{pix}$, pixel pitch, gain, photodetection efficiency PDE, pixel recovery time $\tau_{recovery}$, pixel capacitance $C_{pix}$, quenching resistance $R_q$, difference between turn-on and turn-off voltage for the Geiger avalanche $V_{bd}-V_{off}$, are presented in table~\ref{tab:sipmparameters}\footnote{Note, that before our measurements Zecotek MAPD-3A SiPMs were utilized at NA61 PSD for several years and were already slightly irradiated. $V_{bd}-V_{off}$ was not measured for these samples because they were unable to distinguish single photon peaks which typically happens after irradiation by fluence around $10^{9}$ -- $10^{10}$ n$_{\textrm{eq}}$/cm$^2$~\cite{GARUTTI201969}.}.

\begin{table}[ht]
\centering
\caption{\label{tab:sipmparameters} Parameters of investigated SiPMs produced by various manufacturers. Most of parameters are typical and vary from sample to sample. All SiPMs have $3\times3$~mm$^2$ area. Pixel pitch for Sensl SiPMs is calculated based on number of pixels and SiPM area, so it is bigger than claimed by manufacturer. Values of $C_{pix}$, $R_{q}$ and $V_{bd}-V_{off}$ are from our measurements.}
\smallskip
\resizebox{\textwidth}{!}{%
\begin{tabular}{|c|c|c|c|c|c|c|c|c|}
\hline
 &\multicolumn{2}{|c|}{Zecotek MAPD}&\multicolumn{2}{|c|}{Hamamatsu MPPC}&\multicolumn{2}{|c|}{Ketek SiPM PM33}&\multicolumn{2}{|c|}{Sensl SiPM uF}\\
 \cline{2-9}
 &3 A&3 N&S12572&S14160&15&50&C 30020&B 30020\\
 &&&-010P&-1310PS&-WB-A0&&&\\
 &&&old&new&&&&\\
\hline
$V_{bd}$, V&64&88&67&38&27&23&25&25\\
$N_{pix}$&135000&135000&90000&90000&3600&38800&11000&11000\\
Pitch, $\mu$m&8&8&10&10&15&50&29&29\\
Gain&6$\times$$10^{4}$&$10^{5}$&$10^{5}$&$10^{5}$&3$\times$$10^{5}$&6$\times$$10^{6}$&$10^{6}$&$10^{6}$\\
PDE, \%&20&30&10&18&22&40&24&24\\
$\tau_{recovery}$, ns&2000&10000&10&10&13&2000&100&100\\
$C_{pix}$, fF&1.5&1.2&3.2&6.4&19.5&280&63&63\\
$R_{q}$, M$\Omega$&2.7&2.7&2.7&1.6&0.74&0.42&0.4&0.48\\
$V_{bd}-V_{off}$, V&--&0.43&1.7&0.97&0.72&0.15&0.15&0.15\\
\hline
\end{tabular}
}
\end{table}

SiPMs were irradiated at the cyclotron of NPI \v{R}e\v{z} with a "white" (from thermal up to 34 MeV~\cite{Stefanik2014306}) and a quasi mono-energetic (peak at 22 MeV, with thermal neutron background~\cite{Majerle2016139}) neutron spectra and total fluences in the range of $5\times10^{10}$ -- $6\times10^{12}$ n$_{\textrm{eq}}$/cm$^2$. We estimated the fluence values by the activation foil method, gold foils were irradiated together with SiPMs. Fluence was further recalculated to 1MeV equivalent with damage factors k = 1.54 and 1.62 achieved for "white" and a quasi mono-energetic spectra, respectively. Fluence uncertainty is about 15\% due to complicated neutron spectrum~\cite{Majerle2016139,Stefanik2014306}.

\begin{figure}[ht]
%\centering % \begin{center}/\end{center} takes some additional vertical space
\begin{subfigure}{.47\textwidth}
\includegraphics[width=\linewidth]{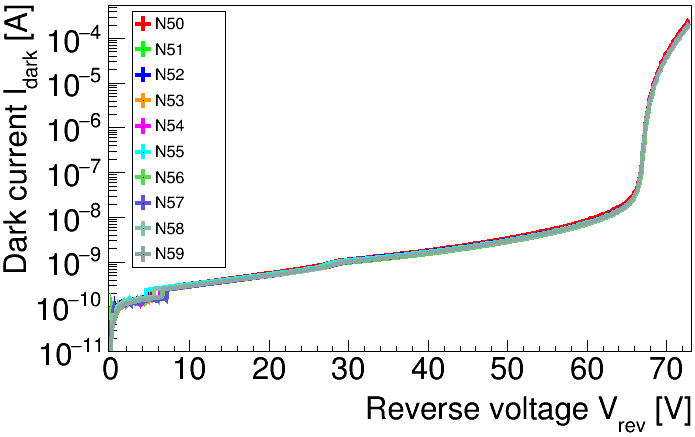}
\caption{}
\end{subfigure}
\qquad
\begin{subfigure}{.47\textwidth}
\includegraphics[width=\linewidth]{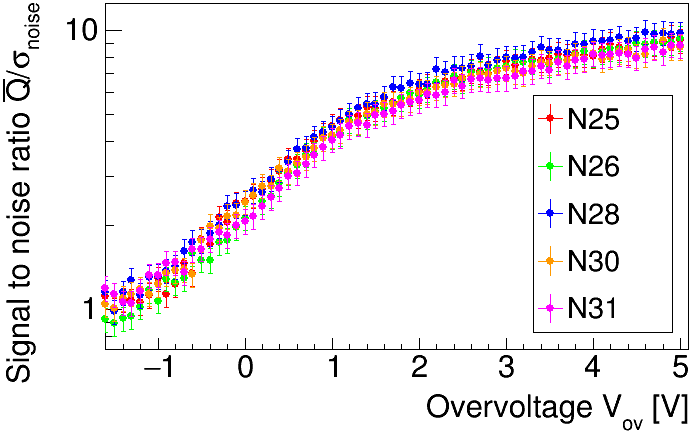}
\caption{}
\end{subfigure}
% "\includegraphics" from the "graphicx" permits to crop (trim+clip)
% and rotate (angle) and image (and much more)
\caption{\label{fig:variability} Dark current variability for ten Hamamatsu MPPCs S12572-010P irradiated by $2.5\times10^{11}$ n$_{\textrm{eq}}$/cm$^2$ (a). Signal to noise ratio variability for five Hamamatsu MPPCs S12572-010P irradiated by $1.6\times10^{12}$ n$_{\textrm{eq}}$/cm$^2$ (b).}
\end{figure}

Samples were irradiated and measured at temperature about 25~$^\circ$C, covered from light. Measurements of SiPMs were performed after several months after irradiation, so self-annealing is considered to be finished.
Laboratory measurement setup includes Keithley 6517A Electrometer, Hioki 3532-50 LCR HiTester, Rohde\&Schwarz RTO1024 Oscilloscope, custom amplifier with gain of 120 and fast 400 nm LED driven by custom pulse generator. 
Dedicated software was developed in NI LabWindows/CVI to automate the measurements.
Voltage step was set to 0.1~V to achieve optimal measurement accuracy. Uncertainties of measured parameters for samples irradiated by different neutron fluences are dominated by sample-to-sample variability. Variability for SiPMs of same type irradiated with the same fluence was typically about 15~\% for dark current and 10~\% for LED response measurements which confirms the uniformity of the sample irradiation, see figure~\ref{fig:variability} in the end of section~\ref{sec:LEDresponse}. Measurement variability for quenching resistance was typically about 8\%, for capacitance it was about 1~\%. 
%appendix~\ref{sec:appendix}
% No significant self-annealing related effects were observed for samples that were remeasured again after several additional months of storage. 

\section{Dark current measurements}

Measurements of SiPM dark currents in reverse bias mode were used to extract the breakdown voltages as maximum of $1/I_{dark}\cdot (dI_{dark}/dV_{rev})$. Extra measurements were performed under illumination for non-irradiated SiPMs to increase the precision of $V_{bd}$ determination. For irradiated SiPMs $V_{bd}$ determined with and without light do not differ. Change of breakdown voltage after irradiation did not exceed 0.5~V.
%, see table [x] in the conclusion. 
Note that $V_{bd}$ is the turn-on voltage for the Geiger avalanche and it differs from the turn-off voltage $V_{off}$~\cite{KLANNER201936}. The latter was measured with help of single photon spectra measurements for non-irradiated SiPMs and the typical difference $V_{bd}-V_{off}$ is presented in table~\ref{tab:sipmparameters}. Unfortunately, after irradiation all the SiPMs lost the ability to distinguish single photons, so $V_{off}$ could not be measured anymore. 

\begin{figure}[!ht]
%\centering % \begin{center}/\end{center} takes some additional vertical space
\begin{subfigure}{.47\textwidth}
\includegraphics[width=\linewidth]{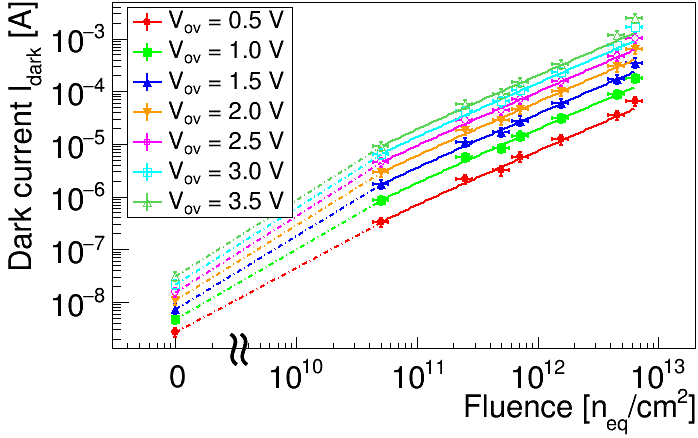}
\caption{}
\end{subfigure}
\qquad
\begin{subfigure}{.47\textwidth}
\includegraphics[width=\linewidth]{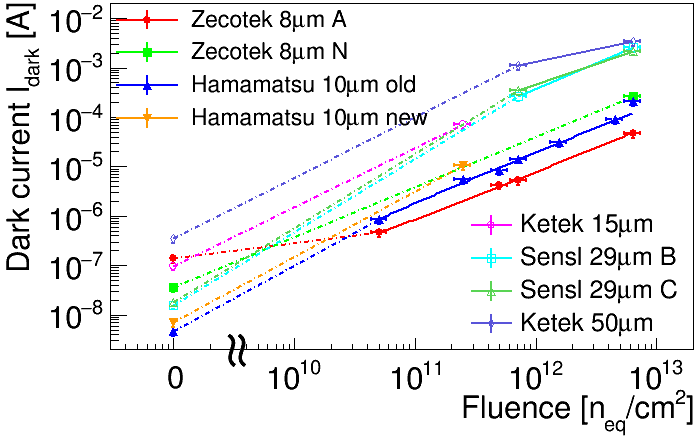}
\caption{}
\end{subfigure}
% "\includegraphics" from the "graphicx" permits to crop (trim+clip)
% and rotate (angle) and image (and much more)
\caption{\label{fig:Idark_Vrev} Dependence of dark current on fluence for Hamamatsu MPPC S12572-010P at different overvoltages (a) and for all the investigated SiPMs at overvoltage $V_{OV}$ = 1 V (b).}
\end{figure}

Figure~\ref{fig:Idark_Vrev} presents dependence of dark current on fluence observed for the investigated SiPMs at different values of overvoltage $V_{OV} = V_{rev} - V_{bd}$. Comparison of different SiPMs is provided for $V_{OV}$ = 1 V because highly irradiated Ketek and Sensl SiPMs reach 10 mA limit of electrometer right after 1 V~\cite{MIKHAYLOV2018NIMA}.
All the SiPMs follow the trend of linear dark current increase with fluence which is typically observed for silicon sensors. Dark current increased in up to 5 orders of magnitude for highly irradiated SiPMs which resulted in huge noise and power consumption making them hardly applicable, see section~\ref{sec:calorperf}. Analysed data suggest that value of dark current after irradiation\footnote{Generally, data before the irradiation are not so straight forward to interpret. Note, that Zecotek MAPD-3A has quite high dark current before irradiation because before our measurements these samples were utilized at NA61 PSD for several years and were already slightly irradiated.} directly depend on the pixel size, i.e. the bigger the pixels -- the higher the dark current.

\begin{figure}[ht]
%\centering % \begin{center}/\end{center} takes some additional vertical space
\begin{subfigure}{.47\textwidth}
\includegraphics[width=\linewidth]{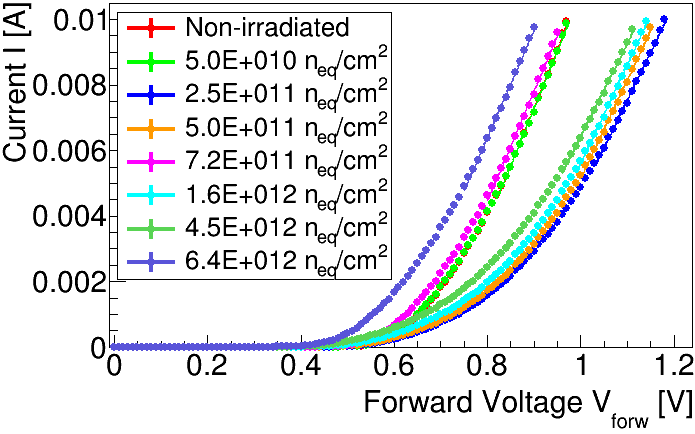}
\caption{}
\end{subfigure}
\qquad
\begin{subfigure}{.47\textwidth}
\includegraphics[width=\linewidth]{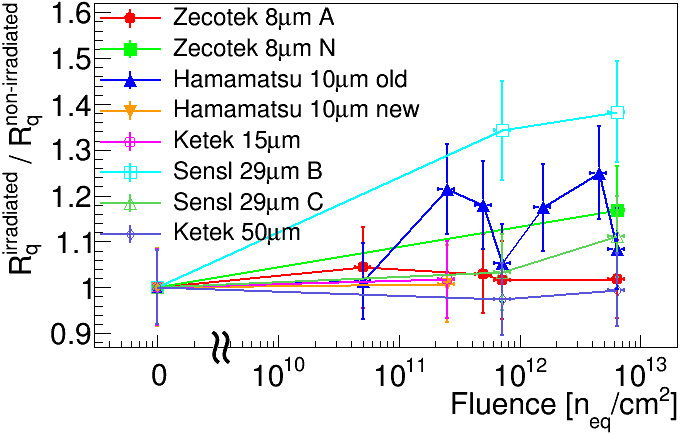}
\caption{}
\end{subfigure}
% "\includegraphics" from the "graphicx" permits to crop (trim+clip)
% and rotate (angle) and image (and much more)
\caption{\label{fig:Idark_Vforw} Dependence of dark current on forward voltage for Hamamatsu MPPC S12572-010P for different fluences (a). Ratio of quenching resistance after/before irradiation for all the investigated SiPMs (b).}
\end{figure}

Measurements of dark current versus forward voltage are exemplified in figure~\ref{fig:Idark_Vforw}~(a). SiPM quenching resistances were extracted from these measurements as $R_q \approx N_{pix}/(dI_{dark}/dV_{forw})$~\cite{KLANNER201936}. Linear fit was performed at the very end of $I_{dark}(V_{forw})$ curve due to its deviation  from linearity in the lower range. Changes of SiPM quenching resistance after the irradiation by $\Phi$ < $10^{11}$ n$_{\textrm{eq}}$/cm$^2$ are below the uncertainty of 8 \% as shown in figure~\ref{fig:Idark_Vforw}~(b). For higher fluences $R_q$ seems to increase by up to 20 \% for some SiPMs\footnote{Only single sample of Sensl uF-B30200 was measured before the irradiation and it exhibited quite strange dark current dependence which could explain the higher deviation of $R_q$ after irradiation}. Absolute values of $R_q$ before irradiation can be found in table~\ref{tab:sipmparameters}.

\section{Capacitance measurements}
\label{sec:capacitance}

\begin{figure}[ht]
%\centering % \begin{center}/\end{center} takes some additional vertical space
\begin{subfigure}{.47\textwidth}
\includegraphics[width=\linewidth]{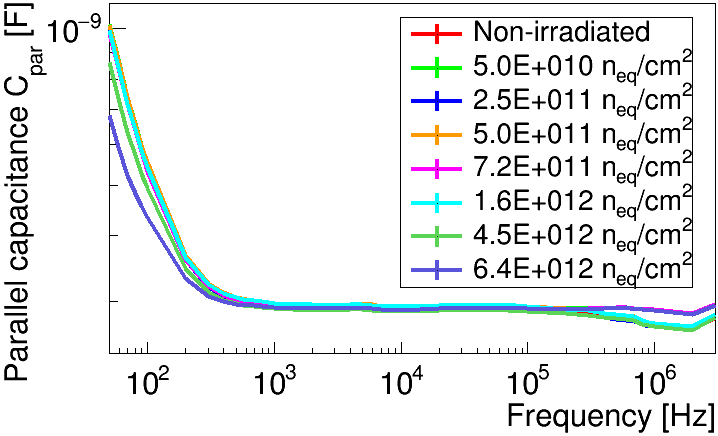}
\caption{}
\end{subfigure}
\qquad
\begin{subfigure}{.47\textwidth}
\includegraphics[width=\linewidth]{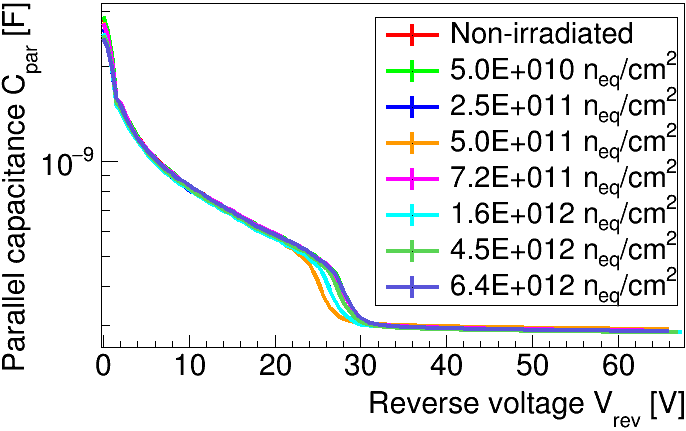}
\caption{}
\end{subfigure}
%\par\medskip
\centering
\begin{subfigure}{.47\textwidth}
\includegraphics[width=\linewidth]{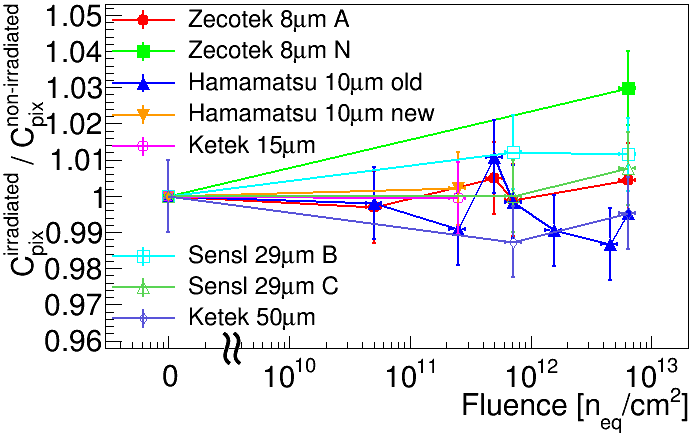}
\caption{}
\end{subfigure}
% "\includegraphics" from the "graphicx" permits to crop (trim+clip)
% and rotate (angle) and image (and much more)
\caption{\label{fig:Capacitance} Dependence of capacitance on test signal frequency below the breakdown for Hamamatsu MPPC S12572-010P for different fluences (a). Dependence of capacitance on reverse voltage at 10 kHz for Hamamatsu MPPC S12572-010P for different fluences (b). Ratio of pixel capacitance after/before irradiation for all the investigated SiPMs (c).}
\end{figure}

Capacitance-frequency measurements of all SiPMs before and after irradiation biased at voltage $0.9V_{bd}$ were carried out in parallel equivalent circuit mode, for example see figure~\ref{fig:Capacitance}~(a). 
Based on these results, stable intermediate frequency of 10 kHz was chosen for further investigation. Pixel capacitance was calculated from parallel capacitance as $C_{pix} \approx C_{par}/N_{pix}$, which is valid for intermediate frequencies where parasitic and quenching capacitances are negligible~\cite{KLANNER201936}. We are interested in $C_{pix}$ above the depletion voltage $V_{dep}$ which is visible around 30 V for Hamamatsu SiPMs in figure~\ref{fig:Capacitance}~(b). However, for Ketek and Sensl SiPMs full depletion is not reached until the breakdown voltage and capacitance measurement in the SiPM avalanche region is not a straight forward task. We assume that the change of $C_{par}$ above $V_{bd}$ is not too big and determine $C_{pix}$ at 1~V below $V_{bd}$. 

Figure~\ref{fig:Capacitance}~(c) shows that pixel capacitances did not change after the irradiation for most SiPMs\footnote{Only for Zecotek MAPD-3N $C_{pix}$ increased by 3 \% which is higher than the typical uncertainty of 1~\%.  However, this uncertainty is based on variability of other samples, while only single irradiated and single non-irradiated Zecotek MAPD-3N samples were investigated, therefore their variability might be higher.}. Absolute values of $C_{pix}$ before irradiation are presented in table~\ref{tab:sipmparameters}. 
SiPM gain is defined as $G(V_{rev}) \approx (C_{pix} + C_q)\cdot (V_{rev} - V_{off})/q_0$, where quenching capacitance $C_q$ is typically an order of magnitude lower than $C_{pix}$~\cite{KLANNER201936}. We might assume that $V_{off}$ only slightly change after irradiation similarly to $V_{bd}$. Then we can conclude that SiPM gain did not change after irradiation as well as $C_{pix}$.

\section{LED response measurements}
\label{sec:LEDresponse}

\begin{figure}[ht]
%\centering % \begin{center}/\end{center} takes some additional vertical space
\begin{subfigure}{.47\textwidth}
\includegraphics[width=\linewidth]{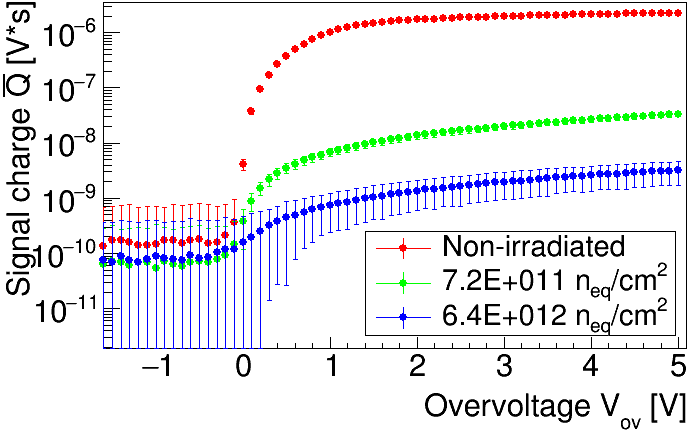}
\caption{}
\end{subfigure}
\qquad
\begin{subfigure}{.47\textwidth}
\includegraphics[width=\linewidth]{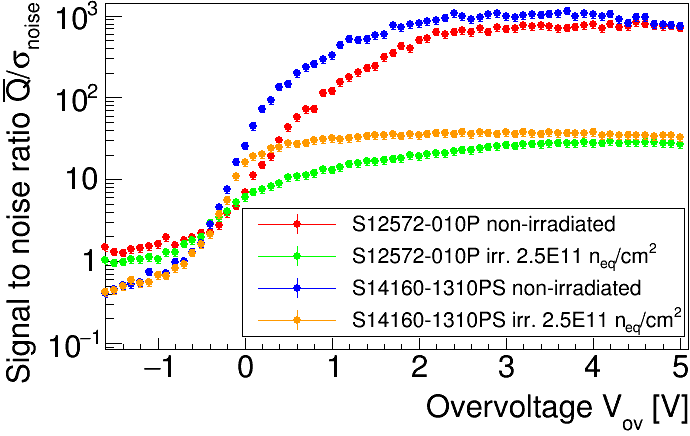}
\caption{}
\end{subfigure}
% "\includegraphics" from the "graphicx" permits to crop (trim+clip)
% and rotate (angle) and image (and much more)
\caption{\label{fig:LEDsignals} Dependence of LED response on overvoltage for Ketek SiPM PM-3350 before and after irradiation (a). Standard deviation of LED response $\sigma_{Q}$ is used as uncertainty here to visualise a poor signal visibility for the most irradiated sample. Dependence of signal to noise ratio before and after irradiation for Hamamatsu 10~$\mu$m SiPMs of old S12572-010P and new S14160-1310PS version (b).}
\end{figure}

The most important is to measure SiPM's ability to serve as a photodetector after the irradiation. For this purpose response of SiPM to LED pulses with constant amplitude and 10~ns width was measured versus overvoltage. Measured signal charge $\bar{Q}$ is defined as mean of the integral area measurement with window gate of 100~ns. Noise estimation is based on standard deviation $\sigma_{Noise}$ of the measured integral without the light exposure. Both are  expressed in nV$\cdot$s. Signal to noise ratio is defined as $SNR = \bar{Q}/\sigma_{Noise}$ and signal resolution is defined as $Res_{Q} = \sigma_{Q}/\bar{Q}$. Signal amplitude was chosen so that non-irradiated and the most irradiated sample could still detect it. For the least radiation hard samples such as Ketek PM-3350 with the largest $50\times50~\mu$m$^2$ pixels we had to choose such a big signal that non-irradiated sample would saturate almost immediately after the breakdown as shown in figure~\ref{fig:LEDsignals}~(a). Due to this fact ratios of signal parameters are presented at overvoltage of 1 V. For lower overvoltages SiPM response is extremely dependent on variation of $V_{bd}$, also it can be too small for highly irradiated samples. 
Hamamatsu SiPMs after irradiation with $\Phi$ = $2.5\times 10^{11}$ n$_{\textrm{eq}}$/cm$^2$ exhibit signal to noise ratio above 10 which is considered to be sufficient for the calorimeter operation, see figure~\ref{fig:LEDsignals}~(b). 

\begin{figure}[ht]
%\centering % \begin{center}/\end{center} takes some additional vertical space
\begin{subfigure}{.47\textwidth}
\includegraphics[width=\linewidth]{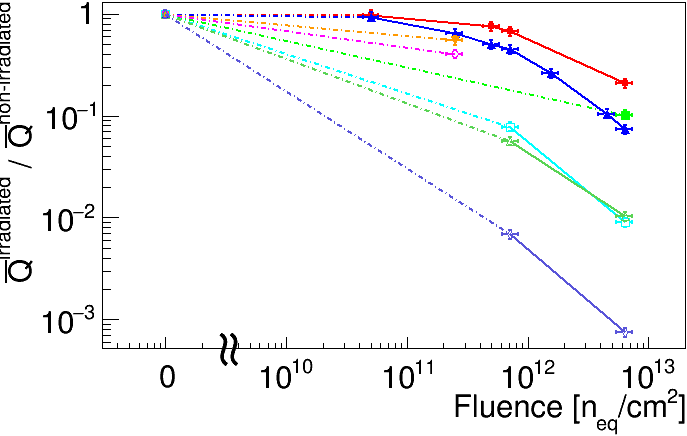}
\caption{}
\end{subfigure}
\qquad
\begin{subfigure}{.47\textwidth}
\includegraphics[width=\linewidth]{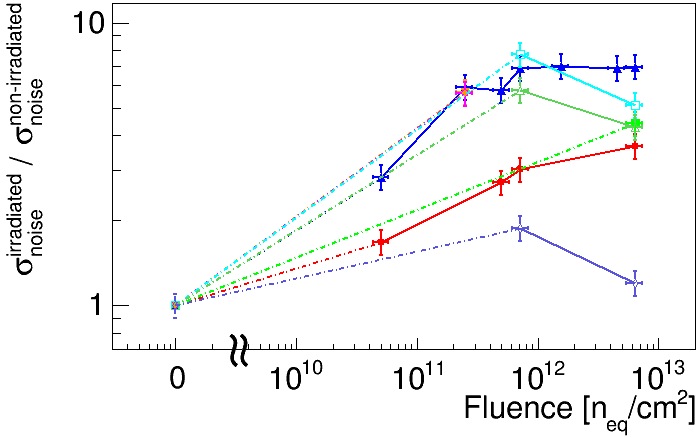}
\caption{}
\end{subfigure}
%\par\bigskip
\begin{subfigure}{.47\textwidth}
\includegraphics[width=\linewidth]{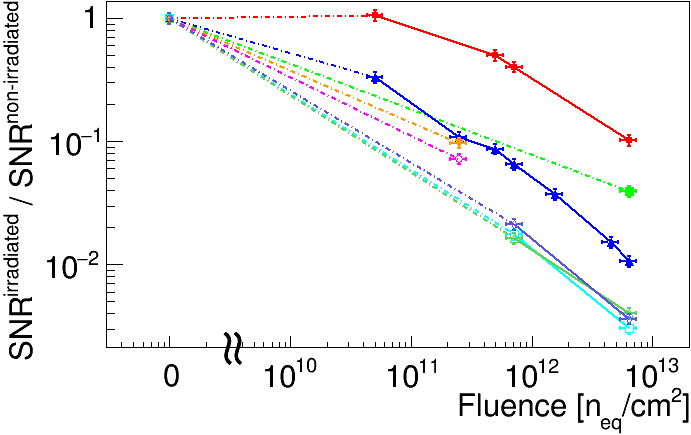}
\caption{}
\end{subfigure}
\qquad
\begin{subfigure}{.47\textwidth}
\includegraphics[width=\linewidth]{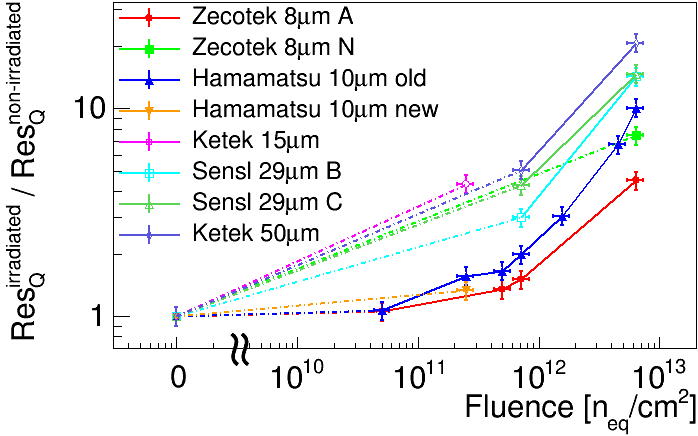}
\caption{}
\end{subfigure}
\caption{\label{fig:LEDratios} Ratios of LED response (a), noise (b), SNR (c) and resolution (d) at overvoltage $V_{OV}$~=~1~V before/after irradiation for all the investigated SiPMs. Note, that drop of signal parameter ratios represent a lower limit for SiPMs with pixel pith > 20 $\mu$m because they exhibit partially saturated signal already at $V_{OV} = $ 1~V, see figure~\ref{fig:LEDsignals}~(a) for example.}
\end{figure}

Drastic degradation of SiPMs' response to LED and consequently signal to noise ratio up to 3 orders of magnitude was observed after irradiation by $\Phi$ = $6.4\times 10^{12}$ n$_{\textrm{eq}}$/cm$^2$, see figure~\ref{fig:LEDratios}~(a, c). No signal degradation was observed for $\Phi$ < $2.5\times 10^{11}$ n$_{\textrm{eq}}$/cm$^2$. 
SiPM signal response to incoming light is essentially a convolution of gain, photodetection efficiency PDE and excess charge factor $ECF$. $ECF$ is responsible for production of secondary correlated Geiger discharges and it is typically~$\leq$~1.2~\cite{KLANNER201936}. Most likely, PDE decreased after irradiation due to change of internal structure of SiPM and/or due to individual pixel failures. Alternatively, SiPM gain could have decreased if indirect gain assessment based on pixel capacitance measurement presented in section~\ref{sec:capacitance} is not valid for irradiated SiPMs. Unfortunately, there is no direct way to measure SiPM gain and PDE after irradiation. 
SiPM noise increased in only 2~--~10 times for $\Phi$ < $10^{12}$ n$_{\textrm{eq}}$/cm$^2$ and then saturated or even decreased, see figure~\ref{fig:LEDratios}~(b). Presented relative decrease of SiPM noise shall be regarded as the lower limit because noise of non-irradiated SiPMs at $V_{OV} = $~1~V is mostly produced by amplification and readout circuitry. 
Noise decrease could be explained with the decrease of SiPM gain which compensate for noise added by higher dark current. 
SiPM resolution degraded by more than factor 10 as shown in figure ~\ref{fig:LEDratios}~(d). 
Relative degradation is smaller for resolution than for SNR because
standard deviation of LED response $\sigma_{Q}$ before irradiation is more than 10 times higher than standard deviation of noise $\sigma_{Noise}$ but after irradiation they become comparable\footnote{$\sigma_{Q}$ depends on Poisson statistics of light counting, so it scales with the measured charge $\bar{Q}$ which is very big before irradiation. Noise on contrary does not depend on $\bar{Q}$ and can be very small before irradiation. After irradiation $\bar{Q}$ become very small and $\sigma_{Q}$ become dominated by $\sigma_{Noise}$, so both become comparable.
%For example, at  $V_{OV}$~=~1~V old Hamamatsu SiPMs before irradiation had $SNR \approx 2.4E-8/1.4e-10 \sim 170$ and $Res_{Q} \approx 1.7E-9/2.4E-8*100\% \sim 7\%$, after the highest irradiation it became $SNR \approx 1.1E-9/9.9e-10 \sim 1.1$ and $Res_{Q} \approx 9.7e-10/1.1E-9*100\% \sim 88\%$.
}.
Similarly to dark currents, SiPM LED response and SNR after irradiation is directly dependent on the pixel size, i.e. the bigger the pixels -- the more the SiPM response degrade with accumulated fluence.

\section{PSD calorimeter performance}
\label{sec:calorperf}

PSD single module response to proton beams was studied in the momentum range of 2 -- 80 GeV/c, including tests at CERN PS beamline for 2 -- 10 GeV/c and tests at CERN NA61 beamline for 10 -- 80 GeV/c.
Module was consequently equipped with 3 batches of Hamamatsu MPPCs S12572-010P irradiated by $2.5\times10^{11}$, $1.6\times10^{12}$ and $4.5\times10^{12}$ n$_{\textrm{eq}}$/cm$^2$. These SiPMs were chosen for the superior radiation hardness with respect to SiPMs produced by Ketek and Sensl manufacturers. We assume that this is largely due to small pixel size of $10\times10~\mu$m$^2$ of Hamamatsu SiPMs. Moreover, small pixel size is important for us as it increases the calorimeter dynamic range. One may note that Zecotek MAPDs have even smaller pixels and better radiation hardness. Unfortunately, due to very large pixel recovery time about several microseconds they cannot withstand the high event rates up to 1~MHz and cannot be utilized at our detectors~\cite{MIKHAYLOV2018NIMA}.

\begin{figure}[ht]
%\centering % \begin{center}/\end{center} takes some additional vertical space
\begin{subfigure}{.47\textwidth}
\includegraphics[width=\linewidth]{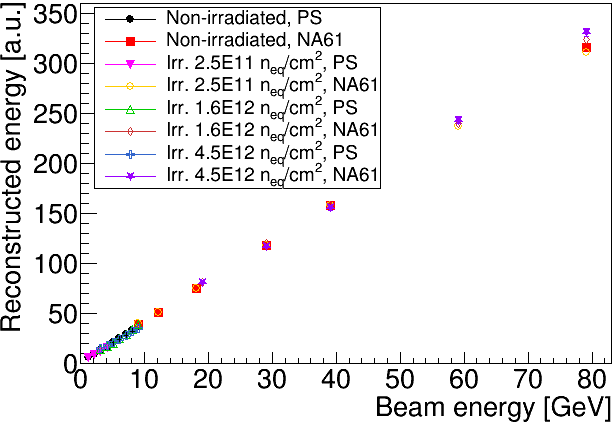}
\caption{}
\end{subfigure}
\qquad
\begin{subfigure}{.47\textwidth}
\includegraphics[width=\linewidth]{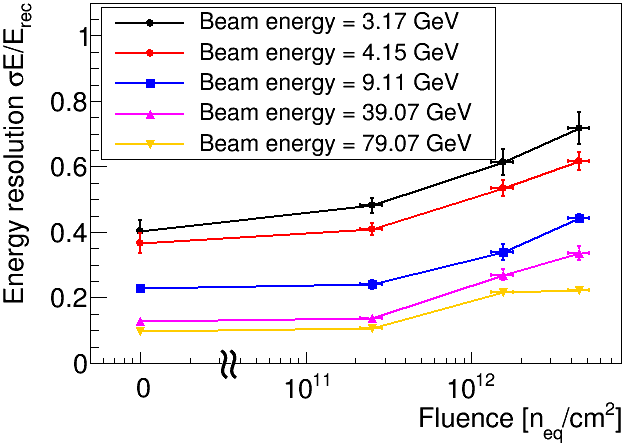}
\caption{}
\end{subfigure}
% "\includegraphics" from the "graphicx" permits to crop (trim+clip)
% and rotate (angle) and image (and much more)
\caption{\label{fig:PSDradperf} Single module proton energy linearity vs beam energy (a). Resolution vs fluence for different beam energies (b). SiPMs were operated at overvoltage = 3V and room temperature. Supply voltage of each SiPM was tuned online with respect to measured temperature to keep $V_{bd}$ and gain stable during the operation.}
\end{figure}

Figure~\ref{fig:PSDradperf}~(a) shows that linearity of the calorimeter response did not suffer from the radiation. Figure~\ref{fig:PSDradperf}~(b) presents the degradation of energy resolution with accumulated fluence. Only slight deterioration is observed after irradiation by $2.5\times10^{11}$ n$_{\textrm{eq}}$/cm$^2$ which is the worst case scenario for a one year of experiment operation\footnote{Results presented for higher fluences of $1.6\times10^{12}$ and $4.5\times10^{12}$~n$_{\textrm{eq}}$/cm$^2$ are an upper limit of resolution degradation because only first 5 sections of the module were equipped with SiPMs and external voltage supply was used due to high SiPM power consumption.}. Modular detector structure of our fix-target heavy-ion collision experiments allows to exchange the most damaged SiPMs every year if necessary.

% Important notes:
% We suggest not to abbreviate: ``section'', ``appendix'', ``figure'' and ``table'', but ``eq.'' and ``ref.'' are welcome. Also, please do not use \texttt{\textbackslash emph} or \texttt{\textbackslash it} for latin abbreviaitons: i.e., et al., e.g., vs., etc.
% For internal references use label-refs: see section~\ref{sec:intro}.
% Bibliographic citations can be done with cite: refs.~\cite{a,b,c}.
% See figure~\ref{fig:i} and table~\ref{tab:i}.

\section{Results discussion and conclusion}

Achieved data suggest only minor changes of SiPM breakdown voltage, quenching resistance, pixel capacitance and gain after irradiation which agrees well with investigations from other authors summarized in~\cite{GARUTTI201969}. 
We also observed linear dependence of SiPM dark current on neutron fluence 
%in range $5\times10^{10}$ -- 
up to $6\times10^{12}$ n$_{\textrm{eq}}$/cm$^2$, same trend was presented for lower fluences up to $6\times10^{9}$ n$_{\textrm{eq}}$/cm$^2$ in~\cite{Qiang2013, Andreotti_2014}.

We found out that SiPMs' response to LED and signal to noise ratio decrease by 10 -- 1000 times at $\Phi$ > $10^{12}$ n$_{\textrm{eq}}$/cm$^2$ for SiPMs with different pixel sizes. 
%Surprisingly, research data on SiPM light response measurements after high neutron irradiation are limited. 
Musienko et al. measured SNR $\approx$ 5 -- 10 for FBK SiPMs with 10 -- 12.5$~\mu$m pixel pitch that were irradiated by $2\times 10^{12}$ n$_{\textrm{eq}}$/cm$^2$ which is similar to our results for SiPMs with 10$~\mu$m pixel pitch.
Several authors~\cite{Cerioli2019, CALVI2019243, Tsang_2016} observed that dark current and LED response performance of highly irradiated SiPM is greatly improved when it is cooled down to -30~$^\circ$C. Even full recovery of PDE and single photon detection can be accomplished at cryogenic temperatures about 80~K for SiPMs irradiated up to by $10^{14}$ n$_{\textrm{eq}}$/cm$^2$~\cite{CALVI2019243, Tsang_2016}. However, no cooling is planned to be utilized at our hadron calorimeters. 

Both dark current and LED response of irradiated SiPMs scale with the pixel size, namely the smaller the pixels -- the better the radiation hardness. Such a scaling was already observed for SiPM dark currents in~\cite{Andreotti_2014, Heering2008,
Barbosa2012} and for LED response in~\cite{Heering2008}.
SiPM pixel miniaturisation is beneficial because it lowers pixel capacitance resulting in the shorter charge collection time and higher collection efficiency contributing to the PDE improvement. It also generally reduces the sensor deadtime and therefore increases the total dynamic range for the high frequency signals. Decreased active volume and surface of a single pixel results in the decrease of generated dark current, and lowered probability of both the after-pulsing and the cross-talk. 
This suggests the use of SiPMs with the smallest available pixels for the harsh radiation environment.

PSD calorimeter performance with older version of Hamamatsu SiPMs irradiated by $2.5\times10^{11}$~n$_{\textrm{eq}}$/cm$^2$ decreased only slightly. New version was proven to perform in a similar manner which suggests its use at new calorimeters which are now being assembled. 
%More info on the presented research can be found in PhD thesis of V.~Mikhaylov which is currently in preparation.

%\appendix
%\section{Examples of SiPM variability}
%\label{sec:appendix}
%Please always give a title also for appendices.

\acknowledgments

The authors thank the CBM and NA61 collaborations for their support in the tests and thank CERN staff M.~Jeckel and L.~Gatignon for their help in test preparation at the beamlines.
The authors thank the NPI cyclotron and neutron generators team, especially M.~\v{S}tef\'anik and M.~Majerle, for excellent beam conditions and help with irradiation tests that were carried out at the CANAM infrastructure.
We also thank V. Ladygin for provision of Ketek SiPMs and Z. Sadygov for provision of Zekotek SiPMs.
This work was supported by Czech MEYS project no. LM2015049 and EU OP VVV - CZ.02.1.01/0.0/0.0/16\_013/0001677 grant. This work was also partially supported by the Ministry of Science and Higher Education of the Russian Federation, grant N 3.3380.2017/4.7, and by the National Research Nuclear University MEPhI in the framework of the Russian Academic Excellence Project (contract No. 02.a03.21.0005, 27.08.2013).

% \paragraph{Note added.} This is also a good position for notes added after the paper has been written.

%\section*{References}
% We suggest to always provide author, title and journal data:
% in short all the informations that clearly identify a document.

\end{document}